\documentclass{autart} 

\usepackage{amssymb}
\usepackage{amsmath}
\usepackage{amsfonts}                                
\usepackage{enumerate}
\usepackage{mathrsfs}
\usepackage{soul,xcolor}
\usepackage{tabulary}
\usepackage{cite}
\usepackage{graphicx}          
\usepackage{color}      
\usepackage{epsfig} 
\usepackage{epstopdf}
\usepackage{times} 
\usepackage{wrapfig}
\def\qedp{\hspace*{\fill}~{\tiny $\blacksquare$}}
\def\qed{\relax\ifmmode\hskip2em \Box\else\unskip\nobreak\hskip1em $\Box$\fi}

\setstcolor{red}
\usepackage{soul}

\newlength{\bibitemsep}
\newlength{\bibparskip}
\let\oldthebibliography\thebibliography
\renewcommand\thebibliography[1]{%
	\oldthebibliography{#1}%
	\setlength{\parskip}{\bibitemsep}%
	\setlength{\itemsep}{\bibparskip}%
}

\def\qedp{\hspace*{\fill}~{\tiny $\blacksquare$}}
\def\qed{\relax\ifmmode\hskip2em \Box\else\unskip\nobreak\hskip1em $\Box$\fi}

\newtheorem{theorem}{Theorem}
\newtheorem{itlemma}{Lemma}
\newtheorem{itdefinition}{Definition}
\newtheorem{itproposition}{Proposition}
\newtheorem{itresult}{Result}
\newtheorem{itremark}{Remark}
\newtheorem{itassumption}{Assumption}
\newtheorem{itcorollary}{Corollary}
\newtheorem{itexample}{Example}

\newenvironment{proposition}{\begin{itproposition}\rm}{\end{itproposition}}
\newenvironment{remark}{\begin{itremark}\rm}{\end{itremark}}

\newenvironment{lemma}{\begin{itlemma}\rm}{\end{itlemma}}

\newcommand{\efrac}[2]{%
	\mathchoice
	{\ooalign{%
			$\genfrac{}{}{1.8pt}0{\hphantom{#1}}{\hphantom{#2}}$\cr%
			$\color{white}\genfrac{}{}{.6pt}0{\color{black}#1}{\color{black}#2}$}}%
	{\ooalign{%
			$\genfrac{}{}{1.8pt}1{\hphantom{#1}}{\hphantom{#2}}$\cr%
			$\color{white}\genfrac{}{}{.6pt}1{\color{black}#1}{\color{black}#2}$}}%
	{\ooalign{%
			$\genfrac{}{}{1.8pt}2{\hphantom{#1}}{\hphantom{#2}}$\cr%
			$\color{white}\genfrac{}{}{.6pt}2{\color{black}#1}{\color{black}#2}$}}%
	{\ooalign{%
			$\genfrac{}{}{1.8pt}3{\hphantom{#1}}{\hphantom{#2}}$\cr%
			$\color{white}\genfrac{}{}{.6pt}3{\color{black}#1}{\color{black}#2}$}}%
}

\newcommand{\Efrac}[2]{%
	\mathchoice
	{\ooalign{%
			$\genfrac{}{}{1.8pt}0{#1}{#2}$\cr%
			$\color{white}\genfrac{}{}{.6pt}0{\phantom{#1}}{\phantom{#2}}$}}%
	{\ooalign{%
			$\genfrac{}{}{1.8pt}1{#1}{#2}$\cr%
			$\color{white}\genfrac{}{}{.6pt}1{\phantom{#1}}{\phantom{#2}}$}}%
	{\ooalign{%
			$\genfrac{}{}{1.8pt}2{#1}{#2}$\cr%
			$\color{white}\genfrac{}{}{.6pt}2{\phantom{#1}}{\phantom{#2}}$}}%
	{\ooalign{%
			$\genfrac{}{}{1.8pt}3{#1}{#2}$\cr%
			$\color{white}\genfrac{}{}{.6pt}3{\phantom{#1}}{\phantom{#2}}$}}%
}

\AtBeginDocument{
	\addtolength{\abovedisplayskip}{-2ex} 
	\addtolength{\abovedisplayshortskip}{5ex}
	\addtolength{\belowdisplayskip}{-2ex}
	\addtolength{\belowdisplayshortskip}{5ex}
}

\begin{document}

	\begin{frontmatter}
		
		\title{The bottleneck and ceiling effects in quantized tracking control of heterogeneous multi-agent systems under DoS attacks  } 
				\vspace{-3mm}
		
		\vspace{-7mm}

		\author[First]{Shuai Feng}\ead{s.feng@njust.edu.cn},    
		\author[Second]{Maopeng Ran}\ead{ranmp@buaa.edu.cn},               
			\author[First]{Baoyong Zhang}\ead{baoyongzhang@njust.edu.cn},  
		\author[Third]{Lihua Xie}\ead{ELHXIE@ntu.edu.sg},  
		\author[First]{Shengyuan Xu$^*$}\ead{syxu@njust.edu.cn}
						\vspace{-1mm}
		\address[First]{School of Automation, Nanjing University of Science and Technology, Nanjing 210094, China}  				\vspace{-1mm}                                             
		\address[Second]{School of Automation Science and Electrical Engineering, Beihang University, Beijing 100191, China and the Zhongguancun Laboratory, Beijing 100094, China}            
				\vspace{-1mm}
		\address[Third]{School of Electrical and Electronics Engineering, Nanyang Technological University, Singapore 68000, Singapore}


\begin{abstract}                          
In this paper, we investigate tracking control of heterogeneous multi-agent systems under Denial-of-Service (DoS) attacks and state quantization. Dynamic quantized mechanisms are designed for inter-follower communication and leader-follower communication. Zooming-in and out factors, and data rates of both mechanisms for preventing quantizer saturation are provided. Our results show that by tuning the inter-follower quantized controller, one cannot improve the resilience beyond a level determined by the data rate of leader-follower quantized communication, i.e., the ceiling effect. Otherwise, overflow of followers' state quantizer can occur. On the other hand, if one selects a ``large" data rate for leader-follower quantized communication, then the inter-follower quantized communication determines the resilience, and further increasing the data rate for leader-follower quantized communication cannot improve the resilience, i.e., the bottleneck effect. Simulation examples are provided to justify the results of our paper.
\vspace{-2mm}	
\end{abstract}
\vspace{-8mm}
\end{frontmatter}

\section{Introduction}
Control of multi-agent systems has attracted substantial attention of researchers. An agent refers to a subsystem in a large-scale system consisting of multiple agents and can be a different subject depending on the context. For example, an agent is a drone in a UAV swarm or a car in a vehicle platoon \cite{FB-LNS}.
Agents are spatially distributed and hence the realization of control significantly depends on the quality of information exchanged among the agents via wireless communication networks. 
However, the challenges of malicious cyber attacks on information also emerge \cite{cardenas2009challenges}.


The control problems under packet losses have been well studied, e.g. in \cite{you2010minimum} for stochastic packet losses, and the case of DoS attacks inducing malicious packet losses has been studied in \cite{de2015input, lu2017input, pierron2020tree}. This paper deals with DoS attacks. The communication failures induced by DoS can exhibit a temporal profile quite different from those caused by genuine packet losses due to 
network congestion; particularly packet dropouts resulting from malicious DoS need not follow a given class of probability distributions \cite{sastry}, and therefore the analysis techniques relying on probabilistic arguments may not be applicable. 

Attention has been focused on centralized systems under DoS attacks for achieving stability\cite{feng2020tac, 8353464}, attack detection and control \cite{bhowmick2020availability} and state estimation\cite{ding2018attacks}. 
In \cite{HSF-SM:16-siam}, the authors study a stabilization problem under energy-constrained PWM DoS signals, and propose time and event triggering strategies under partially known and unknown attack information, respectively.
The paper \cite{befekadu2015risk} describes DoS attacks by a Markov modulated model, in which the control packets are jammed stochastically. In \cite{gupta2016dynamic}, the authors model the interplay between a DoS attacker and a defender as a two-player zero-sum game and compute the saddle-point equilibrium.
Recently, multi-agent systems under DoS attacks have been increasingly investigated, e.g., consensus, state estimation and real-time attack detection \cite{xu2019distributed, amini2020rq, biron2018real}. 
In \cite{deng2022resilient}, the authors propose a 
novel data-driven based algorithm to estimate the unknown switching matrices of the exosystems, and realize output regulation under DoS for heterogeneous multi-agent systems.

For multi-agent systems, the amount of data can be large especially when the number of agents is large. Consequently, agents can be subject to bandwidth limitation, and hence signals are subject to coarse quantization \cite{ishii2002limited, kashyap2007quantized, dibaji2017resilient,rikos2022non, XARGAY2014841,kar2009distributed}. 
Dynamic quantization for centralized systems in the presence of DoS attacks has been recently studied in \cite{feng2020tac, liu2021resilient}, in which the quantization mechanisms have zooming-out and in capabilities for mitigating the influence of DoS-induced packet losses and achieving asymptotic stability, respectively.
Dynamic quantization has also been established for consensus of multi-agent systems without packet losses\cite{you2011network}, in which only the zooming-in capability is developed.
Recently, the papers \cite{feng2020arxiv, feng2023tcns, ran2022quantized} study consensus problems
under DoS attacks and dynamic quantization. In \cite{feng2020arxiv, feng2023tcns}, quantization mechanisms with zooming-in and out capabilities are developed for homogeneous agents. An approach for designing a tight zooming-out parameter is provided, but the condition for consensus is subject to an additional constraint of DoS frequency\cite{feng2023tcns}. As will be shown later, the approach in our paper is free from the additional DoS-frequency constraint.

In practice, a variety of agents may need to cooperate to complete a complicated task, i.e., cooperation of heterogeneous multi-agent systems \cite{li2021cooperative, ding2019distributed}. Consequently, those results for the control of homogeneous agents may not hold. In \cite{ran2022quantized}, heterogeneous nonlinear multi-agent systems under DoS and quantization were studied, while the focus is on the common consensus problem instead of tracking control. 
In this paper, we are interested in output tracking control of heterogeneous multi-agent systems under quantized communication and DoS attacks, in which the quantizer has a finite quantization range. 
The quantized controllers to be designed should prevent quantizer saturation all the time. Unlike DoS-free scenarios, the tracking errors can diverge under DoS attacks, and consequently some measurements can overflow the range of the quantizer if the quantized controllers for DoS-free scenarios are implemented.



For quantized tracking control of multi-agent systems, one can classify the overall information flow into the part for leader-follower quantized communication and the part for inter-follower quantized communication. In \cite{ran2021practical, feng2020arxiv}, the leader-follower communication and inter-follower communication adopt an identical quantization mechanism. However,  this may not provide optimal performance in communication resource allocation and control. One may also lose the insights about the interactions between the leader-follower and inter-follower quantized communication.
For example, in order to save communication resource, one can select a data rate for the leader-follower quantized communication as close as possible to the minimum data rate \cite{1310461}, under which the followers can still estimate the leader's state asymptotically. As will be shown in this paper, such a data rate (of the leader-follower quantized communication) can influence the inter-follower quantized communication, i.e., the quantizer for quantizing followers' state can have overflow problem. In this paper, resilience refers to a bound charactering the amount of DoS attacks under which multi-agent systems can achieve quantized tracking control.
Under DoS attacks, it is interesting and also important to know whether the leader-follower quantized communication or the inter-follower quantized communication influences more on the resilience. From a practical viewpoint, this can actually lead us to the optimal strategies of limited bandwidth allocation for real applications.


The paper's contributions are twofold: a) design heterogeneous dynamic quantized controllers for the tracking control of heterogeneous multi-agent systems, b) discover the interactions between the leader-follower quantized communication and inter-follower quantized communication, and the consequent influences on the resilience of tracking control against DoS attacks. 
We reveal that 
the leader-follower quantized communication can cause ceiling effect and the inter-follower communication can cause bottleneck effect. Specifically, for a given communication topology and a number of bits for the quantization of the leader's state, by tuning the followers' quantized controller, one cannot improve the resilience beyond a level determined by the data rate of leader-follower quantized communication, i.e., the ceiling effect. Otherwise, overflow problems for quantizing followers' state can occur.   
For the bottleneck effect, if one selects a ``large" data rate for leader-follower quantized communication, further enlarging it cannot improve the resilience. Then the inter-follower quantized communication is the bottleneck of the resilience, which depends on the choices of the zooming-in and out factors for followers' state quantization. We emphasize that, by our quantized controller design,  the tightness for zooming-in factor is the same for DoS-free case \cite{you2011network}, and the zooming-out factor is tight, i.e., arbitrarily approaching the spectral radius of the leader's dynamic matrix.

This paper is organized as follows. In Section 2, we introduce the framework consisting of the control objectives, the heterogeneous leader and followers, and the class of DoS attacks. Section 3 presents the design of quantized controllers. The results of the paper are presented in Section 4. 
Numerical examples are presented in Section 5, and finally Section 6 ends the paper with conclusions and future research.

\textbf{Notation.} We denote by  $\mathbb R$ the set of reals. Given $b \in \mathbb R$, $\mathbb R_{\geq b}$ and $\mathbb R_{>b}$ denote the sets of reals no smaller than $b$ and reals greater than $b$, respectively; $\mathbb R_{\le b}$ and $\mathbb R_{<b}$ represent the sets of reals no larger than $b$ and reals smaller than $b$, respectively; $\mathbb Z$ denotes the set of integers. For any $c \in \mathbb Z$, we denote $\mathbb Z_{\ge c} := \{c,c+ 1,\cdots\}$. Let $\lfloor v \rfloor$ be the floor function such that $\lfloor v \rfloor= \max\{o\in \mathbb{Z}|o\le v\}$. 
Given a vector $y$ and a matrix $\Gamma$, let $\|y\|$ and $\|y\|_\infty$ denote the $2 $- and $\infty$- norms of vector $y$, respectively, and $\|\Gamma\|$ and $\|\Gamma\|_\infty$ represent the corresponding induced norms of matrix $\Gamma$. Let $\rho(\Gamma)$ denote the spectral radius of $\Gamma$. Given an interval $\mathcal{I}$, $|\mathcal{I}|$ denotes its length. The Kronecker product is denoted by $\otimes$. We let ``$\efrac{\,\cdot\,}{\,\cdot\,}$" represent element-wise division of vectors, e.g., $\Efrac{[x_1\,\, x_2]^T}{[y_1\,\, y_2]^T}=[x_1/y_1 \,\, x_2/y_2]^T$. 
Let ``$\circ$" denote the element-wise multiplication of two vectors, e.g., $[a_1\,\,a_2]^T \circ [b_1\,\,b_2]^T =[a_1b_1\,\, a_2 b_2]^T$.
In our paper, a switched system refers to a class of systems in the form:
$
	x(k+1)= A_i x(k), i \in \{1, 2, \cdots, c\}, 
$
in which the value of $i$ depends on the switching rules.

\section{Framework}

\textbf{Communication graph:}
We let graph $\mathcal{G} = (\mathcal{N},\mathcal{E})$ denote the communication topology among the followers, where $\mathcal{N}=\{1, 2, \cdots, N \}$ denotes the set of follower agents and $\mathcal{E} \subseteq \mathcal{N} \times \mathcal{N}$ denotes the set of edges. Let $\mathcal N_i$ denote the set of the neighbors of agent $i$, where $i=1, 2, \cdots, N$. We assume that the graph $\mathcal{G}$ is undirected and connected. Let $\mathcal A _ {\mathcal G}= [a_{ij}] \in \mathbb{R}^{N\times N} $ denote the adjacency matrix of $\mathcal {G}$, where $a_{ij} > 0$ if and only if $j \in \mathcal N_i$ and $a_{ii}=0$. Define the Laplacian matrix $\mathcal L_ G = [l_{ij}]  \in \mathbb{R}^{N\times N} $, in which $l_{ii} = \sum_{j = 1 }^{N} a_{ij}$ and $l_{ij} = - a_{ij} $ if $i \ne j $. For the leader-follower communication, we assume that only a fraction of the followers can receive information from the leader. Let $a_{i0}$ represent the leader-follower interaction, i.e., if follower $i$ can directly receive the information from the leader, then $a_{i0}>0$, and otherwise $a_{i0}=0$. Moreover, we let the diagonal matrix be $ \mathcal D= \text{diag}(a_{10}, a_{20}, \cdots, a_{N0}) \in \mathbb{R}^{N \times N}$. Let $\tilde \lambda_i$ ($i=1, 2, \cdots, N$) denote the eigenvalues of $\mathcal L_ G + \mathcal D $.

\textbf{System description:} The follower agents interacting over the network $\mathcal{G}$ are expressed as heterogeneous systems:
\begin{subequations}\label{system}
	\begin{align}
	x_i(k+1) &= A_i x_i(k) +B_i u_i(k) \\
	y_i(k)&= C_i x_i(k)
	\end{align}
\end{subequations}
where $i \in \mathcal N$, $x_i(k)\in \mathbb{R}^{n_i }$ denotes the state of agent $i$, $u_i(k) \in \mathbb{R}^{w_i}$ denotes its control input, $y_i(k)\in \mathbb{R}^{v}$ denotes its output, and $A_i \in \mathbb R^{n_i \times n_i}$, $B_i \in \mathbb R^{n_i \times w_i}$ and $C_i\in \mathbb R^{v\times n_i}$. The sampling time between $k$ and $k+1$ is $\Delta$.
We assume that $(A_i,B_i)$ is stabilizable and thus there exists a feedback gain $K_i \in \mathbb R ^{w_i \times n_i}$ such that $\rho(A_i+B_i K_i)<1$. We assume that an upper bound of the initial condition $x_i(0)$ is known, i.e., $\| x_i(0)\|_\infty \le C_{x_0} \in \mathbb{R}_{>0}$ ($i \in \mathcal N$). 
Note that $C_{x_0}$ can be arbitrarily large as long as it satisfies this bound. This is for preventing the overflow of state quantization for the initial condition.
The dynamics of the leader can be described by
\begin{align}\label{leader}
v(k+1) = S v(k)
\end{align}
where $v(k) \in \mathbb R ^{n_v}$ is the state and $S \in \mathbb R^{n_v\times n_v}$. We assume that $\rho(S) \ge 1$. Otherwise, as $v(k) \to 0$ when $k\to \infty$, one can simply realize the tracking control by locally stabilizing (\ref{system}) under which $y_i(k)\to 0$. Similarly, we assume that an upper bound of the initial condition $v(0)$ is known, i.e., $\| v(0)\|_\infty \le C_{v_0} \in \mathbb{R}_{>0}$. 

We assume that $x_i(k)$ is available to the local controller of agent $i$. We also assume that only some of the followers can receive information from the leader.
Followers can exchange information with their neighbor followers. We assume that transmissions are acknowledgment based and free of delay. Note that in a quantized control problem without packet losses, acknowledgments are not necessary \cite{1310461,nair2004stabilizability}. When packet losses and quantization coexist, acknowledgments are needed to ensure the synchronization of the signals in the encoders and decoders  \cite{you2010minimum, wakaiki2019stabilization}. Due to the data rate limitation, all the information exchanged via communication is encoded by limited numbers of bits. This implies that the transmitted signals are subject to quantization effect. In this paper, we develop two quantization mechanisms: one for followers' state quantization, and another for leader's state quantization.
The motivations of adopting heterogeneous quantizers will be presented later (see Remark \ref{remark 3}).   

\textbf{Quantizer for followers' state:}  Let $\chi \in \mathbb{R}$ be a scalar before quantization and $q_{R_f}(\cdot)$ be the quantization function for scalar
input values as
\begin{align}\setlength{\arraycolsep}{3pt}  \label{quantizer}
\!\!q_{R_f} (\chi)\!\! = \!\! 
\left\{\!\!
\begin{array}{lll}
0 & -\sigma < \chi < \sigma & \\
2\psi \sigma & (2\psi-1)\sigma \le \chi \! < \!(2\psi+1)\sigma  \\
2R_f\sigma & \chi \ge  (2R_f+1) \sigma&         \\
-q_{R_f} (-\chi) & \chi \le - \sigma & 
\end{array}
\right.
\end{align}
where $R_f\in \mathbb{Z}_{>0}$ is to be designed and $\psi =1, 2, \cdots, R_f$, and $\sigma \in \mathbb{R}_{>0}$. If the quantizer is unsaturated such that $|\chi| \le (2R_f+1)\sigma $, then the  quantization error satisfies 
$
|\chi - q_{R_f}(\chi)| \le \sigma.
$

\textbf{Quantizer for leader's state:}
 Let 
$
\pi_l:= e_l/\omega_l
$
be the $l$-th signal in a vector before quantization and
$q_{\mathcal R_l} (\pi_l)$ represents the quantized signal of $\pi_l$ encoded by $\mathcal R_l$ bits, where $l= 1, 2, \cdots, v$. The choices of $\mathcal R_l$, $e_l \in \mathbb{R}$ and $\omega_l\in\mathbb{R}_{ > 0}$ will be specified later. We implement a uniform quantizer such that 
\begin{align} \label{uniform quantizer}
q_{\mathcal R_l} (\pi_l) : = 
\left \{ \begin{array}{ll} 
\frac{\lfloor 2^{\mathcal R_l-1} \pi_l \rfloor + 0.5}{ 2^{\mathcal R_l-1} }, &  \quad  \textrm{if } -1 \le \pi_l < 1  \\ 
1- \frac{0.5}{2^{\mathcal R_l-1}}, & \quad  \textrm{if } \pi_l=1
\end{array} \right. 
\end{align} 
if $\mathcal R_l \in \mathbb{Z}_{\ge 1} $, and 
$
q _{\mathcal R_l} (\pi_l) =0
$
if $\mathcal R_l = 0 $. Note that for any $\omega_l \in \mathbb{R}_{> 0} $ the following holds:
\begin{align} \label{Property of quantizer}
\left| e_l - \omega_l q_{\mathcal R_l} \left(e_l / \omega_l \right)  \right| \le \omega_l / 2^{\mathcal R_l}, \quad \textrm{if}   \,|e_l|/ \omega_l \le 1
\end{align}
for both cases \cite{you2010minimum}.

\subsection{Time-constrained DoS}

We refer to DoS as the event under which all the encoded signals cannot be received by the decoders and it affects all the communication links simultaneously.  
We consider a general DoS model
that describes the attacker's action by the frequency of DoS attacks and their duration. Let 
$\{h_q\}_{q \in \mathbb Z_{\ge 0}}$ with $h_0 \geq \Delta$ denote the sequence 
of DoS \emph{off/on} transitions, that is,
the time instants at which DoS exhibits 
a transition from zero (transmissions are possible) to one 
(transmissions are impossible).
Hence,
$
H_q :=\{h_q\} \cup [h_q,h_q+\tau_q[  
$
represents the $q$-th DoS time-interval, of a length $\tau_q \in \mathbb R_{\geq 0}$,
over which the network is in DoS status. If $\tau_q=0$, then
$H_q$ takes the form of a single pulse at $h_q$.  
Given $\tau,t \in \mathbb R_{\geq0}$ with $t\geq\tau$, 
let $n(\tau,t)$
denote the number of DoS \emph{off/on} transitions
over $[\tau,t]$, and let 
$
\Xi(\tau,t) := \bigcup_{q \in \mathbb Z_{\ge 0}} H_q  \, \cap  \, [\tau,t] 
$
be the subset of $[\tau,t]$ where the network is in DoS status. 

\textbf{Assumption 1} \cite{de2015input}
	(\emph{DoS frequency}). 
	There exist constants 
	$\eta \in \mathbb R_{\geq 0}$ and 
	$\tau_D \in \mathbb R_{> 0}$ such that
$
	n(\tau,t)  \, \leq \,  \eta + \frac{t-\tau}{\tau_D}
$
	for all  $\tau,t \in \mathbb R_{\geq \Delta}$ with $t\geq\tau$.
	\qedp

\textbf{Assumption 2} \cite{de2015input}
	(\emph{DoS duration}). 
	There exist constants $\kappa \in \mathbb R_{\geq 0}$ and $T  \in \mathbb R_{>1}$ such that
$
	|\Xi(\tau,t)|  \, \leq \,  \kappa + \frac{t-\tau}{T}
$
	for all  $\tau,t \in \mathbb R_{\geq \Delta}$ with $t\geq\tau$. 
	\qedp

\begin{remark}
	In Assumption 1, $\tau_D$ can be considered as the average dwell-time between 
	consecutive DoS off/on transitions, while $\eta$ is the chattering bound.
	Assumption 2 expresses a similar 
	requirement with respect to the duration of DoS. 
	It expresses the property that, on the average,
	the total duration over which communication is 
	interrupted does not exceed a certain \emph{fraction} of time,
	as specified by $1/T$.
	The constant $\kappa$ plays the role
	of a regularization term. It is needed because
	during a DoS interval, one has $|\Xi(h_q,h_q+\tau_q)| = \tau_q >  \tau_q /T$.
	Thus $\kappa$ serves to make Assumption 2 consistent. 
	Conditions $\tau_D>0$ and $T>1$ imply that DoS cannot occur at an infinitely
	fast rate or be always active. \qedp
\end{remark}

\textbf{Control objectives:} We seek to design a distributed controller $u_i (k)$ such that the quantizers in (\ref{quantizer}) and (\ref{uniform quantizer}) do not saturate under DoS attacks; the outputs of the followers can track the state of the leader:
$
\lim_{k  \to \infty }   y_i(k)  = v(k).
$

\begin{figure}[t]
	\begin{center}
		\includegraphics[width=0.49 \textwidth]{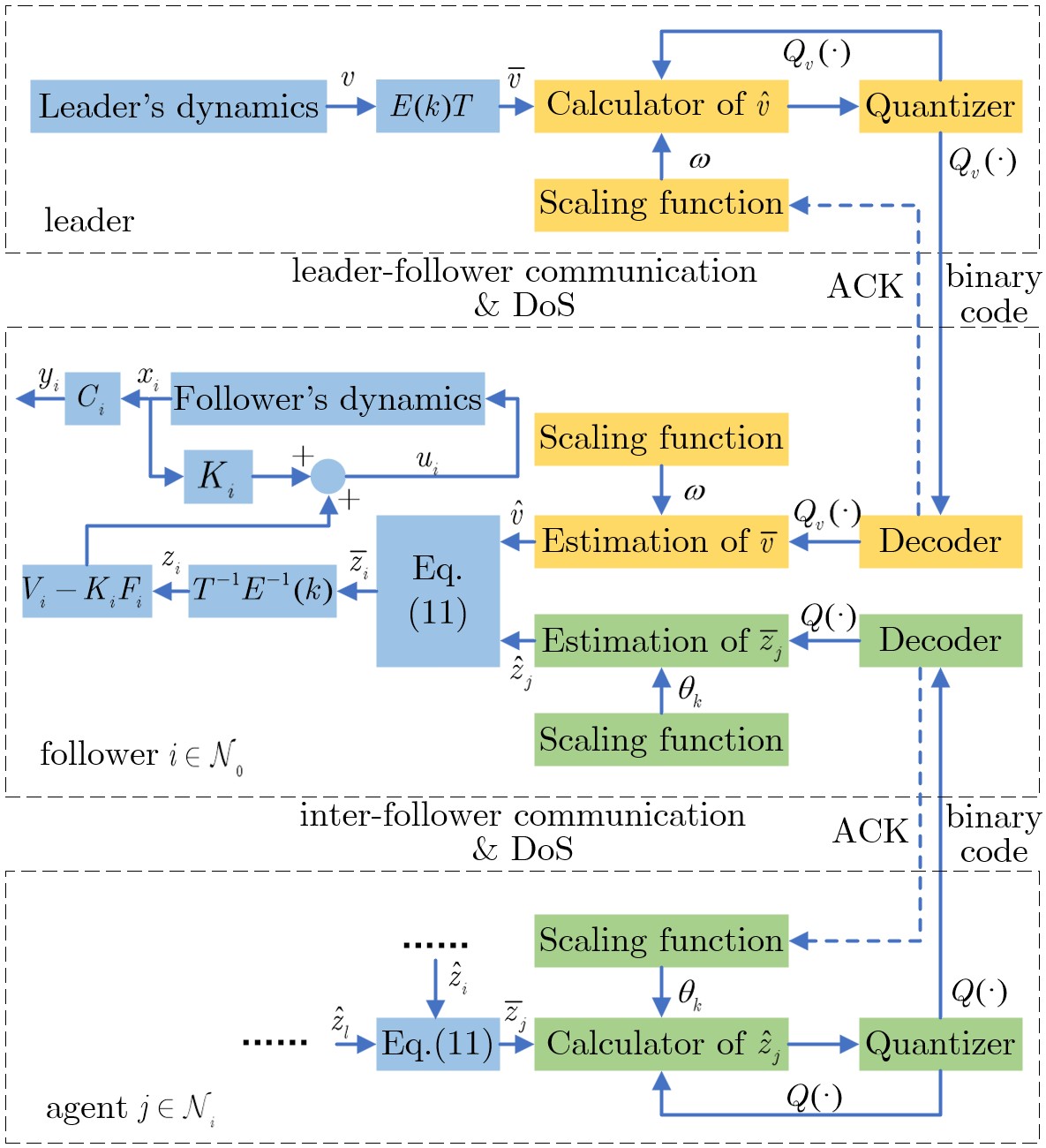} \\
		\vspace{-2mm}
		\linespread{1}\caption{Control structure over the network. Agent $i \in \mathcal N_0$ is a direct follower of the leader. Agent $j\in \mathcal N_i$ is a neighbor of agent $i$. $l$ is the index for agent $l\in \mathcal N_j$. ACK denotes acknowledgments. }\label{fig}
	\end{center}
\end{figure}

\section{Controller design}
The following lemma is necessary to relax the constraints of dwell time of switched systems and obtain a tight data rate  (see Remarks \ref{remark 2} and \ref{remark 3}). 

\begin{lemma}\label{lemma 1}
	\cite{1310461} There exist (bounded) matrices $E(k)\in \mathbb R^{n_v\times n_v}$ and $T\in \mathbb R^{n_v\times n_v}$, and a transformation $\bar v(k) = E(k) T v(k)$, possibly time-varying, transforming (\ref{leader}) into 
	\begin{align} 
\!\!\!	\bar v (k+1) = \bar S  \bar v(k), \, \bar S = \text{diag}\left( \bar S_1,   \cdots,  \bar S_p \right) \in \mathbb{R} ^ {n_v \times n_v}
	\end{align}
	 	where $p\in \mathbb{Z}_{\ge 1}$ denotes the number of sub-matrices in $\bar S$.
	Let $r=1, 2, \cdots, p$, then one has
	\begin{align}\setlength{\arraycolsep}{3pt} \label{second Jordan block}
	\bar S_r = \left[\begin{array}{cccc}
	\lambda_r  &1& & \\ & \lambda_r  & 1 & \\ & & \ddots &1 \\ & & & \lambda_r
	\end{array}\right] \in  \mathbb{R}^{n_r \times n_r}
	\end{align}
	corresponding to the real eigenvalue $\lambda_r \in \mathbb R$ of $S$, and 
	\begin{align}\label{second Jordan block complex}
	&\bar S_r  = \left[
	\!\!\begin{array}{llll}\setlength{\arraycolsep}{5pt} 
	\zeta_r I_2 &r^{-1}(\phi)& & \\ & \zeta_r I_2 & r^{-1}(\phi) & \\ & & \ddots &r^{-1}(\phi) \\ & & & \zeta_r I_2
	\end{array}
	\!\!\right]  \in  \mathbb{R}^{2n_r \times 2n_r}
	\end{align}
	corresponding to the complex eigenvalues $\lambda_r=\zeta_r(\cos \phi \pm i\sin \phi  )$ with $\zeta_r \ge 0$ and
	\begin{align}\setlength{\arraycolsep}{3pt}
	r(\phi)= \left[
	\begin{array}{rr}
	\cos \phi & \sin \phi \\
	-\sin \phi & \cos \phi
	\end{array}\right],\,\,
	I_2= \left[
	\begin{array}{cc}
	1 & 0 \\
	0  & 1  
	\end{array}\right]. \nonumber \quad  \quad \quad  \quad\quad  \,\,\,\, \text{\qedp}
	\end{align}
\end{lemma}

If $S$ has only real eigenvalues, the time-varying part in the transformation $\bar v(k) = E(k) T v(k)$ is not needed, i.e., $\bar v(k) = T v(k)$. Then, one only has the Jordan blocks in (\ref{second Jordan block}).

\textbf{Assumption 3}\,\,	For $i \in \mathcal N$, we assume that there exists $(F_i, V_i)$ satisfying
$
F_i S=A_i F_i  + B_i V_i$ and
$I_v 	= C_i F_i,  
$
where $I_v$ is the identity matrix of dimension $n_v\times n_v$, and $V_i \in \mathbb R ^{w_i \times n_v}$ and $F_i \in \mathbb R ^{n_i \times n_v}$.  \qedp

Assumption 3 is a typical assumption for cooperative control of heterogeneous multi-agent systems \cite{li2021cooperative}. We refer the readers to \cite{su2011cooperative} for more details. 

\textbf{Controller design:} As shown in Figure \ref{fig}, for follower agent $i\in \mathcal N$, we propose the local control input
\begin{align}\label{controller0}
u_i(k) = K_i x_i(k) + (V_i - K_i F_i)  z_i (k)
\end{align}
in which
\begin{align}\label{12}
z_i(k):=T^{-1} E^{-1}(k) \bar z_i (k) \in \mathbb R^{n_v}.
\end{align}
The dynamics of $\bar z_i(k)$ in (\ref{12}) follows
\begin{align}\label{zi}
&\bar z_i(k+1) =  \nonumber \\
&\left\{
\begin{array}{ll}
\bar S \bar z_i(k) + \bar K \sum_{j=1} ^{N} a_{ij} (\hat z_j  (k) - \hat z_i  (k)) \\
\quad \quad\quad  + \, \bar K   a_{i0} (  \hat v  (k) -  \hat z_i(k)), &\text{if}\,k \notin H_q \\
\bar S \bar z_i(k), & \text{if}\, k \in H_q
\end{array}
\right.
\end{align}
where $\hat z_j(k)\in \mathbb R ^{n_v}$ and $\hat v(k)\in \mathbb R ^{n_v}$ are the estimates of $\bar z_j(k)$ and $\bar v(k)$, respectively, and $\bar K  \in \mathbb R^ {n_v \times n_v}$ is a design parameter to be given later. 
The computation of $\hat z_j(k)$ in (\ref{zi}) follows
\begin{align}\setlength{\arraycolsep}{3pt}  \label{eq estimator}
\hat  z_j(k)
\!\!=\!\!
\left\{ \!\!\!
\begin{array}{ll}
\bar S \hat  z_j(k-1) \!+\! \theta_{k-1} Q \left(\!  \frac{  \bar z_j(k) -   \bar S \hat  z_j( k-1 )}{\theta_{k-1}} \!  \right)  & \text{if $k \notin H_q$ } \\
\bar S \hat  z_j( k-1  )             & \text{if $k \in  H_q$} 
\end{array}
\right.   
\end{align}
where $j\in \{i\} \bigcup \mathcal N_i$ and $Q(\cdot)=[q_{R_f} (\cdot)\cdots q_{R_f} (\cdot)]^T$ is the vector version of (\ref{quantizer}).  
The scaling parameter $\theta_k$ updates as
\begin{align}\label{eq h}
\theta_k = 
\left\{
\begin{array}{ll}
\gamma_1 \theta_{k-1}, & \quad \text{if $ k   \notin H_q $} \\
\gamma_2 \theta_{k-1}, & \quad \text{if $ k  \in H_q $   }
\end{array}
\right.
\end{align}
where $\gamma_1<1$ and $\gamma_2 >  1$ are 
zooming in and out parameters,
respectively. We assume that the algorithms in (\ref{eq estimator}) and (\ref{eq h}) are embedded in the encoders and decoders of $i\in \mathcal N$ with identical initial conditions. Under DoS attacks, the variables in $Q(\cdot)$ may diverge. Therefore, the quantizers must zoom out by using $\gamma_2$ and increase their ranges so that the states can be measured properly. If the transmissions succeed, the quantizers zoom in and $\theta_k \in \mathbb R_{>0}$ decreases by using $\gamma_1$. By adjusting the scaling parameter $\theta_{k}$ in $Q(\cdot)$ dynamically, the state will be kept within the limited quantization range without saturation and can converge asymptotically.  
The ranges of $\gamma_1$ and $\gamma_2$, and $\theta_0$ will be specified later. Note that $\gamma_1, \gamma_2$ and $\theta_0$ in our paper are homogeneous among the followers. It is interesting to design distributed scaling parameters, e.g., $\gamma_1 ^i$ and $\gamma_2 ^i$. In this situation, the new scaling parameter $\theta_k$  in (\ref{eq h}) will be a vector, and zooming-in and out factors composed by $\gamma_1^i$ and $\gamma_2^i$ will be matrices. This case will be left for future research.

The update of $\hat v(k)$ in (\ref{zi}) is given by 
\begin{align}\label{Jordan predictor}
\!\!\! \hat v  (k)  \!\! = \!\!
 \left\{\!\! \begin{array}{ll} 
\bar S \Big( \hat v (k-1) \\ 
\,\, - \left. \omega (k-1) \circ Q_v\left(\Efrac{\bar S \hat v(k-1)-\bar v(k)}{   \bar S  \omega(k-1)}\right) \!  \right), & k \notin \!H_q \\
\bar S \hat v (k-1),  & k \in \! H_q \\
\end{array}\right. 
\end{align}
in which $Q_v(\cdot)$ is the vector form of (\ref{uniform quantizer}), $\omega(k) \in \mathbb R ^ {n_v} $ is a scaling vector for quantizing $\bar v(k)$, ``$\circ$" is the element-wise multiplication and ``$\efrac{\,\,\cdot\,\,}{\cdot}$" is the element-wise division. 
The computation of $\omega(k)$ in (\ref{Jordan predictor}) is given as follows:
\begin{align} \label{Jordan case J}
\!\! \!\!\!  
\left \{\!\! \begin{array}{l}
\omega(k) =\left \{ \begin{array}{ll}
\tilde {S}H \omega(k-1), & k \notin H_q \\
\tilde S \omega(k-1), & k \in H_q
\end{array}\right.  \\
H \!=\! \text{diag}(2^{-R_1}I_1, 2^{-R_2}I_2, \cdots, 2^{-R_p}I_p)  \!\in \! \mathbb{R}^{n_v \times n_v}
\end{array}\right.
\end{align} 
%
where the block diagonal matrix $\tilde S = \text{diag}(\tilde S_1, \cdots, \tilde S_p) \in \mathbb R ^{n_v \times n_v}$. If $r$ corresponds to a real eigenvalue $\lambda_r \in \mathbb R$, then $\tilde S_r = \bar S_r$ in (\ref{second Jordan block}). Otherwise, $\tilde S_r$ takes
\begin{align}
\tilde S_r \!\! = \! \!
\left[\begin{array}{cccc}\setlength{\arraycolsep}{2pt} 
\zeta_r I_2 &O & & \\ & \zeta_r I_2 & O & \\ & & \ddots \\ & & &\zeta_r I_2
\end{array}\right] \!\!  \in  \!  \mathbb{R}^{2n_r \times 2n_r},
O \!=\!
\left[\begin{array}{cc} \setlength{\arraycolsep}{3pt}
1& 1 \\
1&1
\end{array}\right]\!\!.
\end{align}
Similar to \cite{feng2020tac}, the parameter $R_r$ ($r=1, \cdots, p$) in $H$ is the number of bits for the quantization process corresponding to $\bar S_r$. If $R_r$ is determined, then $\mathcal R_l$ in (\ref{uniform quantizer}) is determined as well.   
At last, we assume that the initial conditions in the encoding and decoding systems are identical and satisfy
\begin{align} \label{Jordan initial condition}
\hat v_l (0  ) = 0,\,\,  \omega_l(0 ) >  |\bar{v} _l(0 )|,\,\,  l= {1, 2, \cdots, v}
\end{align} 
where $\hat v_l(k) \in \mathbb R$ and $\omega_l(k) \in \mathbb R$ denote the $l$-th element in vectors $\hat v(k)= [\hat v_1(k)\cdots \hat v_l(k) \cdots \hat v_v(k)]^T$ and $\omega(k)= [\omega_1(k)\cdots \omega_l(k) \cdots   \omega_v(k)]^T$, respectively. 

				\vspace{-1mm}
\section{Stability analysis}
				\vspace{-1mm}

Define vectors $\bar z(k) := [\bar z_1 ^T (k)     \cdots   \bar z_N ^T (k)   ]^T$, $\hat z(k) := [\hat z_1 ^T (k)     \cdots   \hat z_N ^T (k) ]^T$ and $\delta(k):=\bar z(k)- \mathbf 1 _N \otimes \bar v(k)$. Define errors $e_z (k):=\bar z(k)-\hat z(k)$ (of estimating $\bar z(k)$) and $e_v(k):= \hat v(k) - \bar v(k)$ (of estimating $\bar v(k)$). Define matrices
$
G:= \bar S _N - (\mathcal L _G + \mathcal D) \otimes \bar K, \bar S _N := I_N \otimes \bar  S, W:=  \mathcal D  \otimes \bar K, 
P:= (\mathcal L_G +  \mathcal D) \otimes \bar K$ and $Z:=  \bar S _N +  (\mathcal L _G + \mathcal D)  \otimes \bar K. 
$
We define $\alpha(k) : = \delta(k)/\theta_k$, $\xi_z(k) := e_z(k)/\theta_k$ and $\xi_v(k) := e_v(k)/\theta_k$. Accordingly, we obtain the dynamics of $\alpha(k)$ and $\xi_z(k)$ in four cases as follows. The dynamics of $\xi_v(k)$ will be analyzed later. 

\textbf{Case I: $k+1\notin H_q$ and $k\notin H_q$}
\begin{subequations}\label{35}
	\begin{align}
	&\!\!\!\! \alpha( k+1 )=  \frac{G}{\gamma_1} \alpha (k) + \frac{P}{\gamma_1}  \xi_z(k)  - \frac{W}{\gamma_1} (\mathbf 1_N \otimes \xi_v(k)) \\
	&\!\!\!\!  \xi_z( k+1 ) = \frac{Z}{\gamma_1} \xi _z (k)  -  \frac{P}{\gamma_1} \alpha(k)   - \frac{W}{\gamma_1} (\mathbf 1_N \otimes \xi_v(k)) \nonumber\\
	& \quad   -    \frac{1}{\gamma_1} Q\left(Z \xi_z(k)  -  P \alpha(k)   - W (\mathbf 1_N \otimes \xi_v(k))\right)  
	\end{align}
\end{subequations}
\textbf{Case II: $k + 1\notin H_q$ and $k\in H_q$}
\begin{subequations}
	\begin{align}
	&\alpha( k+1 )=  \frac{\bar S_N }{\gamma_1} \alpha (k)  \\
	&\xi_z ( k+1 ) = \frac{\bar S_N }{\gamma_1} \xi_z(k)  -  \frac{1}{\gamma_1} Q\left( \bar S_N \xi_z(k) \right) 
	\end{align}	
\end{subequations}
\textbf{Case III: $k+1\in H_q$ and $k\notin H_q$}
\begin{subequations}
	\begin{align}
	&\!\!\!\! \alpha( k+1 )=  \frac{G}{\gamma_2} \alpha (k) + \frac{P}{\gamma_2}  \xi_z(k)  - \frac{W}{\gamma_2} (\mathbf 1_N \otimes \xi_v(k)) \\
	&\!\!\!\! \xi _z ( k+1 ) = \frac{Z}{\gamma_2} \xi_z(k)  -  \frac{P}{\gamma_2} \alpha(k)   - \frac{W}{\gamma_2} (\mathbf 1_N \otimes \xi_v(k)) 
	\end{align}	
\end{subequations}
\textbf{Case IV: $k +1 \in H_q$ and $k\in H_q$}
\begin{subequations}\label{33}
	\begin{align}
	&\alpha( k+1 )=  \frac{\bar S _N }{\gamma_2} \alpha (k)  \\
	&\xi_z ( k+1 ) = \frac{\bar S _N}{\gamma_2} \xi_z(k).  
	\end{align}
\end{subequations}

The next proposition specifies the ranges of $\gamma_1$, $\gamma_2$ and the value of $(2R_f +1)\sigma$ preventing the saturation of (\ref{quantizer}). Its proof with $C_1$ and $C_2$ is provided in the Appendix. 

\begin{proposition}\label{pro 1}
	Suppose there exists a $E_v \in \mathbb R_{>0}$ such that $\|\xi_v(k)\|_ \infty \le E_v $. Choose the zooming-in factor by $ \rho (G)< \gamma_1 <1 $, the zooming-out factor by $\gamma_2 > \rho(S)$ and $\theta_0 \ge C_{x_0}\gamma_1/\sigma$. Under DoS attacks in Assumptions 1 and 2, the quantizer (\ref{quantizer}) is free of saturation if 
	\begin{align}
\quad	&\!\!\! (2R_f+1)\sigma \nonumber\\
	&\!\!\! \ge C_2 \|\bar S_N \|\left(\!\!  \|Z\|\sqrt{Nv}\frac{\sigma}{\gamma_1} \!\! +\!\!  \|P\| C_1 \!\! +  \!\!  \|W\| \sqrt{Nv}E_v   \!\!\right). \quad  \,\,\text{\qedp} \nonumber
	\end{align}
\end{proposition}

\begin{remark}\label{remark 2}
Lemma \ref{lemma 1} is necessary to obtain ``tight" $\gamma_1$ and $\gamma_2$. Otherwise, the matrices on the diagonals of $\bar S_N$ and $\bar G$ (in (\ref{32})) shall be $S$ and $S- \tilde \lambda_i \bar K$, respectively, which are not necessarily upper-triangular. Then, the stabilization of the switched system regulated by $\bar S_N$ and $\bar G$ can be subject to dwell time constraints. Consequently, $\gamma_1$ and $\gamma_2$ can be much more conservative in order to compensate the state ``jumps" of the switched system as in \cite{feng2020arxiv}. Alternatively, if one seeks to obtain a less conservative zooming-out parameter, DoS frequency (regulating the switching frequency) should be upper bounded in order to ensure that the quantizer is not saturated and consensus is achieved \cite{feng2023tcns}. 
\qedp
\end{remark}

In Proposition \ref{pro 1}, by properly selecting $\gamma_1$ and $\gamma_2$, we have provided the value of $(2R_f+1)\sigma$ for the unsaturation of quantizer (\ref{quantizer}), by supposing $\xi_v(k)$ upper bounded. In the following proposition, we provide the conditions to ensure an upper bounded $\xi_v(k)$, namely a finite $E_v$.

\begin{proposition}\label{theorem 1}
	Supposing that $\gamma_2$ is selected as in Proposition \ref{pro 1}, we have the following results:
	\begin{itemize}
		\item[a)] The quantizer for the leader's state in (\ref{uniform quantizer}) is free of overflow for all $k$. 
		\item[b)] If $\gamma_1$ satisfying Proposition \ref{pro 1} is given, then $R_r$ needs to satisfy $2^{R_r} > \zeta_r /\gamma_1$ so that $E_v$ is finite.
		\item[c)] If $R_r > \log_2 \zeta_r$ is pre-given, then $E_v$ is finite if $\gamma_1$ satisfies $\max \{ \rho(\bar G), \zeta_r / 2^{R_r} \} < \gamma_1<1$ with  $\bar G$ in (\ref{32}).  
	\end{itemize}
\end{proposition}
%

\textbf{Proof.} Let $e_{vl}(k)$ denote the $l$-th element of vector $e_{v}(k)$. To prove a), we will show $|e_{vl}  (k)| \le \omega_l(k)$, which implies quantizer unsaturation in view of (\ref{uniform quantizer}). If $k=1 \notin H_q$, then by (\ref{Jordan predictor}) one has 
\begin{align}\label{43}
e_v   (1)= \bar S   \left(   e_v   (0) - \omega(0)  \circ  Q_v \left( \efrac{ e_v   (0)}{\omega (0)} \right)         \right).
\end{align}
Note that $|e_{vl} (0)| \le \omega_l(0)$ implied by (\ref{Jordan initial condition}). Then according to (\ref{Property of quantizer}), element-wise inequality of (\ref{43}) yields 
\begin{align}\label{42}
|e_{vl} (1)| \le |\bar S_l|  H \omega(0) \le  \tilde S_l  H \omega(0)  = \omega_l (1)
\end{align}
in which $|\bar S_l|$ is the $l$-th row of $|\bar S|$, and $\tilde S _l$ is the $l$-th row of $\tilde S$. Here, $|\bar S|$ is a matrix in which the elements are the absolute values of the corresponding elements in $\bar S$. 
If $k=1 \in H_q$, we have $e_v (1) = \bar S e_v(0)$ by (\ref{Jordan predictor}). Following a similar element-wise analysis as in (\ref{42}), one has 
\begin{align}
\!\!\!\! |e_{vl} (1)| \!\le\! |\bar S_l| [|e_{v1}(0)| \cdots |e_{vv}(0)|]^T  \!\le\! \tilde S _l   \omega(0) \!=\! \omega_l(1).
\end{align}
By the above analysis, we have shown that if overflow does not occur at $k=0$, then no matter $k=1$ is corrupted by DoS or not, overflow cannot occur. By induction, one can confirm that overflow does not occur for all $k$.

To prove b) and c), we first analyze the dynamics of  
\begin{align}\label{44b}
\xi_{vl}(k) 
=\frac{e_{vl}(k)}{\omega_l(k)} \frac{\omega_l(k)}{\theta(k)} 
\vspace{-2mm}
\end{align}
where $\xi_{vl}(k) $ is the $l$-th element in $\xi_{v}(k) $.
We have shown that $|e_{vl}(k)|/\omega_l(k) \le 1 $ in a). In order to ensure $ \xi_{vl}(k)$ is bounded, one needs to ensure that $w_l(k)/\theta(k)$ is upper bounded, namely bounded $w(k)/\theta(k)$ in vector form. Let $\bar \omega(k) : = \omega(k)/\theta_k$. In the absence of DoS, by (\ref{eq h}) and (\ref{Jordan case J}), one has
\begin{align}\label{eq 43}
\bar \omega(k+1) =\frac{ \tilde S  H}{\gamma_1} \bar \omega(k).
\end{align}
It is clear that if b) or c) holds, then $\rho(\tilde S H  /\gamma_1)<1$. In the presence of DoS, we have 
\begin{align}\label{44}
\bar \omega(k+1) =\frac{  \tilde S}{\gamma_2}\bar \omega(k)
\end{align}
where $\rho(\tilde S    /\gamma_2) < 1$ since $\gamma_2 > \rho(\bar S)= \rho(\tilde S)$ in Proposition \ref{pro 1}. Because $\tilde S H$ and $\tilde S$ are upper-triangular matrices, by the stabilization of switched systems, one can confirm that $\bar \omega(k)$ is bounded, and hence there exists a positive and finite $E_v$ such that $\|\xi_v(k)\|_\infty  \le E_v$ in view of (\ref{44b}). 
\qedp

\begin{remark}\label{remark 3}
Proposition \ref{theorem 1} shows that due to the design of $\omega(k)$ and $\tilde S$ therein, the quantizer (\ref{uniform quantizer}) is free of overflow for any $R_r$. However, an inappropriate $R_r$ can induce saturation problems to the quantizer (\ref{quantizer}). 
For instance, arbitrarily approaching the minimum data rate for $R_r$ is not enough. Even if some followers can estimate $\bar v(k)$ under the minimum $R_r$ and quantizer (\ref{uniform quantizer}) is not saturated, the quantizer (\ref{quantizer}) may have overflow problem. 
Specifically, suppose that $R_r$ is sufficiently close to the minimum data rate $\log_2 \zeta_r$. Under such a $R_r$, the quantization error of leader's state $e_v$ converges to zero in the absence of DoS attacks. However, such a $R_r$ does not necessarily satisfy $R_r > \log_2 (\zeta_r/\gamma_1)$ in Proposition 2 b). The violation of Proposition 2 b) can make the dynamics of (\ref{eq 43}) unstable and consequently $E_v$ infinitely large. By Proposition \ref{theorem 1}, if $E_v$ is unbounded, any finite-range quantizer (\ref{quantizer}) for followers' state has the overflow problem.
To solve the problem, there are two options. The first option is selecting a larger $R_r$, i.e., satisfying b) in Proposition \ref{theorem 1} for a given $\gamma_1$. As a second option, if one would like to get close to the minimum $R_r$ \cite{1310461}, then a $\gamma_1$ sufficiently close to 1 is needed as indicated in c). 
As will be shown in Theorem \ref{main result}, when $R_r$ approaches the minimum data rate, one gets a poor resilience.

We emphasize that if one adopts an identical quantization mechanism for the leader's and followers' state as in \cite{feng2020arxiv} (as in (\ref{quantizer}), (\ref{eq estimator}) and (\ref{eq h})), it is difficult to make $R_r$ tight, and the gap between $R_r$ and the minimum data rate is invisible. If one adopts an identical quantization mechanism for the leader's and followers' state as in (\ref{uniform quantizer}), (\ref{Jordan predictor}), (\ref{Jordan case J}), then designing the scaling parameter to ensure the followers' state satisfying $|e_l/\omega_l|$ in (\ref{Property of quantizer}) is one of the major challenges, which will be left for future research. 
\qedp
\end{remark}

As a special case, if the dimension of the $\bar S_r$ corresponding to the spectral radius is 1, the result in Proposition \ref{theorem 1} b) can be relaxed to $2^{R_r} \ge \zeta_r /\gamma_1$. Note that if the dimension of the $\bar S_r$ corresponding to the spectral radius is larger than 1, such a relaxation can make (\ref{eq 43}) unstable. Similarly, if the dimension of the $\bar S_r$ corresponding to the spectral radius is 1, one can let $\gamma_2 = \rho(\tilde S)$. 

Note that Proposition \ref{theorem 1} cannot imply tracking control under DoS. The following main result characterizes the system's resilience of tracking control under DoS attacks. 

\begin{theorem}\label{main result}
Suppose that $\gamma_1$ and $R_r$ satisfy b) or c) in Proposition 2, and $\gamma_2$ satisfies Proposition 1. Then, one can achieve the tracking control if DoS attacks satisfy
\begin{align}\label{50}
\frac{1}{T} + \frac{\Delta}{\tau_D} 
&<  1- \frac{\log_2 \gamma_2}{\log_2 \gamma_2 - \log_2 \gamma_1} \nonumber\\
&< 1 -\frac{\log_2 \gamma_2}{\log_2 \gamma_2 - \log_2 \frac{\zeta_r }{2^{R_r}}}.
\vspace{-2mm}
\end{align}
\end{theorem}
\textbf{Proof.} 
For achieving tracking control, it is necessary that $ \delta(k) \to 0$ as $k \to \infty$. This can be obtained by $\theta_k \to 0$ in view of $\delta(k) = \theta_k \alpha(k)$, in which one can confirm that $\|\alpha(k)\|$ is upper bounded by following a similar analysis in the proof of Proposition \ref{pro 1}. 
If DoS attacks satisfy the first inequality in (\ref{50}), then
one has $\theta_k \to 0$ \cite{feng2020arxiv}. By Proposition \ref{theorem 1}, one can infer that $(R_r, \gamma_1)$ satisfies $\gamma_1 > \zeta_r / 2^{R_r}$. Thus, one can obtain the second inequality in (\ref{50}).



Now we show $y(k)\to v(k)$. Let $A:=\text{diag}(A_1, \cdots, A_N)$, and $B, K, F$ and $V$ take the similar form being block diagonal matrices. Let $\bar E(k):=I_N \otimes E(k)$ and $\bar T:=I_N \otimes T$. Then, in view of Assumption 3 and $p(k):= x(k) - F v(k) $, one has  
$
p(k+1)  
 = (A+BK)\, p(k) + B(KF-V)(\mathbf 1_N  \otimes v(k) -   z(k) ) 
= (A+BK)\, p(k)  
\,\,\,+ B(KF-U) \bar  T^{-1} \bar E ^ {-1}(k)(\mathbf 1_N  \otimes \bar v(k) - \bar z(k))
$, 
in which we have shown that $ \bar z(k)  - \mathbf 1_N \otimes \bar v(k)    = \delta(k) \to 0$. Since $\rho(A+BK)<1$ and  $\|\bar T^{-1}  \bar E ^ {-1}(k)\|$ bounded, one can verify that $p(k) \to 0$ as $k \to \infty$, and hence $Cp(k)=Cx(k) - CF v(k) = y(k) - v(k) \to 0$. \qedp

\begin{remark}
\textbf{a)} By (\ref{50}), it is clear that the resilience depends on the leader-follower communication and inter-follower communication. Small $\gamma_1$ and $\gamma_2$, and a large $R_r$ can improve the system's resilience against DoS. Without affecting the main idea of the paper, we let $\gamma_2$ infinitely approach $\rho(S)$. This is a significant improvement compared with \cite{feng2020arxiv, feng2023tcns}. 


\textbf{b)} If one selects a sufficiently large $R_r$ for leader-follower quantized communication, e.g., $\zeta_r / 2^{R_r} < \rho(G) < \gamma_1$, then the quantized communication among the followers is the bottleneck of resilience. In other words, large $\gamma_1$ and $\gamma_2$ lead to poor resilience no matter how large is $R_r$. Mathematically, this can be confirmed by (\ref{50}). We provide an intuitive explanation. Suppose that one chooses a $R_r$ large enough such that 
the second inequity in (\ref{50}) is infinitely close to $1$. This implies that leader's state has almost no quantization effects, and the leader's direct neighbors can obtain the leader's state by only one successful transmission. Subsequently, communication between the leader and its direct followers is not necessary since the followers can perfectly estimate $v(k)$ locally. 
The rest to be done is that the followers who have $v(k)$ should ``deliver" $v(k)$ to the other followers via the follower-follower communication. Then, it is intuitive to see that, now the ability of cooperative control among followers under DoS becomes the bottleneck, i.e., $\gamma_1$ and $\gamma_2$.

\textbf{c)} The quantized communication between the leader and followers is the ``ceiling" of resilience if one selects a small $R_r$, e.g., $\zeta_r / 2^{R_r}$ is close to 1. In this case, tuning the followers' parameters cannot help much because $\gamma_1$ is lower bounded by $\zeta_r / 2^{R_r}$ for preventing quantizer overflow in view of c) in Proposition 2. In oder to prevent further resilience degradation induced by the quantized communication among followers (the first inequality in (\ref{50})), one needs to make $\gamma_1 \to \zeta_r / 2^{R_r} $ from the right. Then, one can verify that the right hand-side of the first inequality in (\ref{50}) approaches that of the second inequality (the ``ceiling").
One can still try to improve $\gamma_1 <  \zeta_r / 2^{R_r}$ such that the first inequality goes beyond the ``ceiling" determined by $R_r$. However, this can cause quantizer overflow problem to (\ref{quantizer}) due to unbounded $E_v$ (see the bottom plot in Fig. \ref{example2}), which violates the first control objective.

\textbf{d)} Following a)-c), if one has selected a $\gamma_2$ close to $\rho(S)$, a small $\gamma_1$ close to $\rho(G)$ and a $R_r> \log_2 \zeta_r /\gamma_1$, then there is not much room for improving the resilience by quantized controller design. This is because $S$ (determining $\gamma_2$) can depend on the inherent properties of the leader, and $G$ (determining $\gamma_1$) depends on $S$ and communication topology. They together determine the bottleneck. One can attempt to find a $\bar K$ minimizing $\rho(G)$, but this is out of the scope of our paper. \qedp



\end{remark}


By the proof of Theorem \ref{main result}, the dynamics of $p(k)$ determines the convergence rate of $y(k)-v(k)$. It is straightforward that $\|\bar z(k)-\mathbf 1_N \otimes \bar v(k)\| = \|\delta(k)\| = \|\theta_k \alpha(k)\| \le \theta_k C$ influences the convergence speed, where $C>0$ denotes an upper bound of $\|\alpha(k)\|$. Now, one can see that the convergence speed mainly depends on $\theta_k$. In view of (\ref{eq h}), a small $\gamma_1$ satisfying $\max \{\rho(\bar  S - \tilde \lambda _i \bar K) \}_{i \in \mathcal N}<\gamma_1<1$ and a small $\gamma_2 > \rho(S)$, and less times of $k \in H_q$ can lead to a ``fast" convergence of $\theta_k$, which essentially depends on the dynamics of $S$, the eigenvalues of $\mathcal{L+D}$ and the number of packet losses induced by DoS attacks. Fast converge rate also requires more data rate. This can be confirmed by Proposition \ref{pro 1} and Proposition \ref{theorem 1} b). Otherwise, quantizer overflow can occur. To maximize the convergence speed, in general, one should improve the communication topology, select a $\bar K$ minimizing $\max \{\rho(\bar  S - \tilde \lambda _i \bar K) \}_{i \in \mathcal N}$, a $\gamma_1$ sufficiently close to $\max \{\rho(\bar  S - \tilde \lambda _i \bar K) \}_{i \in \mathcal N}$, and a large $R_r$ satisfying Proposition \ref{theorem 1}.     

One can almost recover the number of quantization levels for homogeneous multi-agent systems by making the heterogeneous quantization mechanisms in this paper homogeneous by letting $E_v = \sigma/\gamma_1$. By ``almost", we mean that the values can be different due to parameter selections and dwell time constraints in \cite{feng2020arxiv}. Meanwhile, the  inequalities in (\ref{50}) characterizing the resilience recover to that in \cite{feng2020arxiv}.

If the communication network is free of DoS attacks but subject to quantization, the zooming-out process is not needed and the problem recovers to the classic quantized leader-follower consensus. Note that if one applies heterogeneous quantization mechanisms as proposed in this paper, one still needs b) or c) in Proposition \ref{theorem 1} to hold. Otherwise, follower's quantizer can have overflow problem. Our result can also recover to that of quantization-free network under DoS attacks, namely choosing $\gamma_1 = \max \{\rho(\bar S - \tilde \lambda _i \bar K) \}_{i \in \mathcal N}$ and $\gamma_2 = \rho(S)$. However, such $\gamma_1$ and $\gamma_2$ can lead to infinite quantization levels due to infinite $\alpha(k)$ and $E_v$ caused by $\rho(\bar S_N /\gamma_2) = \rho(G/\gamma_1) =1$ (see the Appendix and (\ref{44}), respectively). This is consistent with quantized consensus for homogeneous multi-agent systems under DoS attacks.

\section{Simulation}

			In this section, we present two examples. Example 1 is mainly for the verification of the ceiling and bottleneck effects, and Example 2 is the application of our control scheme to the tracking control of mobile robots.

\textbf{Example 1:}
We consider a multi-agent system with one leader and four followers. The dynamics of the leader and followers is regulated by $S=[1.1052\,\, 0.1105; 0\,\, 1.1052]$, and $A_i=[0 \,\,1; 0\,\, 0]$, $B_i=[1.1052\,\, -0.8895; 0\,\, 1.1052]$ and $C_i=[1 \,\,0.5; 0\,\, 1]$, respectively. 
The interaction between the leader and followers is represented by $\mathcal D=\text{diag}(1, 0, 1, 0)$. The Laplacian matrix of the graph for inter-followers' communication follows that in \cite{feng2023tcns}. The feedback gain in (\ref{controller0}) is selected as $\bar K_i=-0.1I_2$ and the design parameter in (\ref{zi}) is selected as $K=0.4683I_2$. The transmission interval between $k$ and $k+1$ is $\Delta=0.1$s.

\begin{figure}[t]
	\begin{center}
		\includegraphics[width=0.4 \textwidth]{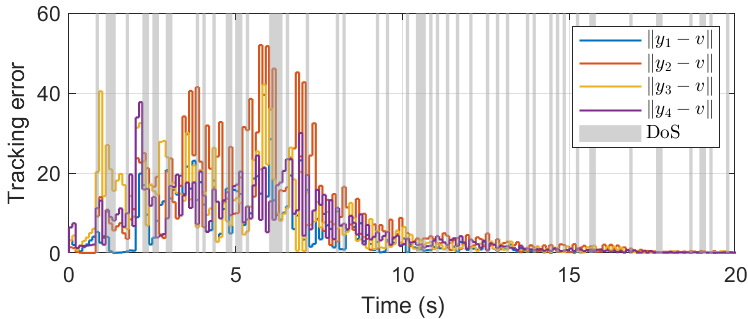} \\
		\includegraphics[width=0.4 \textwidth]{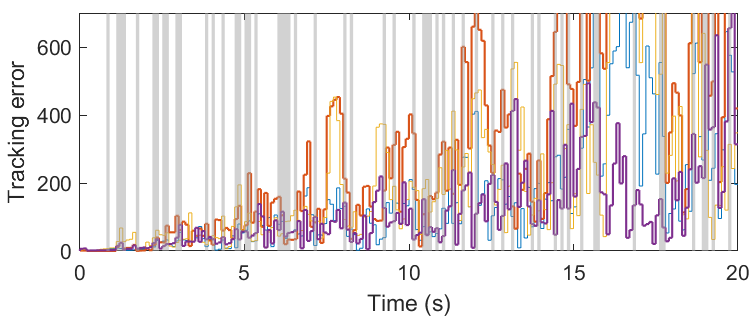} \\
		\vspace{-2mm}
		\linespread{1}\caption{Top: Time response of tracking errors ($\gamma_1=0.92$); Bottom: Time response of tracking errors ($\gamma_1=0.97$) and the legend follows that in the top plot.} \label{example1}
	\end{center}
	\vspace{-2mm}
\end{figure}

One can obtain $\rho(G)=0.9161$ and $\rho(S)=1.1052$. Then we select $\gamma_1=0.92$ and $\gamma_2=1.10521$ following Proposition \ref{pro 1}. We select $R_1=1$ according to Proposition \ref{theorem 1}. By Proposition \ref{pro 1}, we obtain $(2R_f+1)\sigma>7.3638\cdot10^{14}$, which can be encoded by 50 bits. The theoretical sufficient bound of DoS attacks computed by Theorem \ref{main result} is $0.4546$. 
We consider a sustained DoS attack with variable period and duty cycle, generated randomly.
Over a simulation horizon of $20$s, the DoS 
signal yields $|\Xi(0,20)|=5.7$s and $n(0,20)=42$. 
This corresponds to values (averaged over $20$s) 
of $\tau_D\approx 0.4762$ and $1/T \approx 0.2850$,
and hence $\Delta/\tau_D \approx 0.15$ and
$
\Delta/\tau_D + 1/T \approx 0.495
$.
Since our result regarding tolerable DoS attacks is a sufficient condition, one can see from the first plot in Figure \ref{example1} that the tracking control is achieved. We mention that the variables to be quantized in $Q(\cdot)$ in the simulation are between $-7.65$ and $9.2$ (5 bits per state), which is much smaller than the theoretical result, i.e., $  9.2 \ll 7.3638\cdot10^{14} $. The theoretical result is conservative since we have selected $\gamma_1$ and $\gamma_2$ very close to their lower bounds and this can lead to large $C_1$, $C_2$ and $E_v$.

Now we show the bottleneck effect. We select $\gamma_1=0.97$ and let $\gamma_2$ be the same. According to Proposition \ref{theorem 1} b), one sees that $R_1=1$ still satisfies the inequalities. However, for the same DoS pattern studied above, one can see from the second plot in Figure \ref{example1} that the tracking error diverges. 
Note that the value of $R_1$ does not change and hence the ``ceiling" does not change (the second inequality in (\ref{50})). However, the first inequality in (\ref{50}) becomes smaller under a smaller $\gamma_1$, which is actually the bottleneck of resilience causing state divergence. In this example, one still needs 50 bits for encoding a state because one selects a tight $\gamma_2$ for the pair $(\rho(S)/\gamma_2, \rho(G)/\gamma_1)$, and $\rho(S)/\gamma_2$ determines the lower rate for iteration during the computation of $C_1$, $C_2$ and $E_v$. In the simulation, we actually have $-8 \le \|Q (\cdot)\|_\infty \le 8 $ (5 bits per state), though the tracking control fails due to the bottleneck effect.

\begin{figure}[t]
	\begin{center}
		\includegraphics[width=0.4 \textwidth]{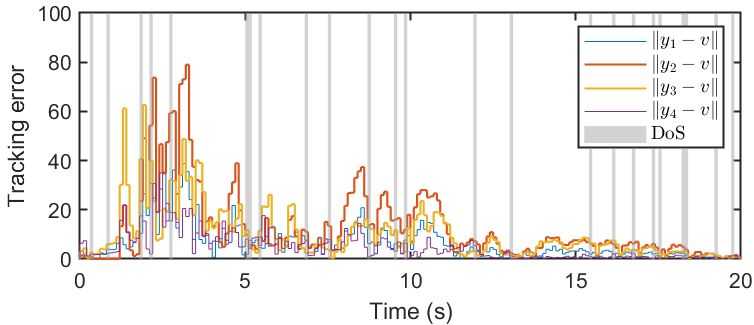} \\
		\includegraphics[width=0.4 \textwidth]{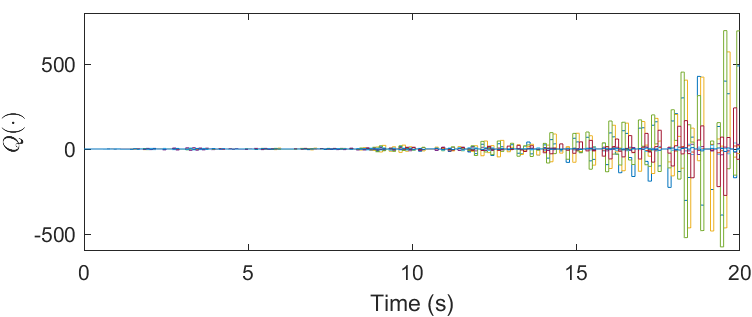} \\
				\vspace{-2mm}
		\linespread{1}\caption{Top: Time response of tracking errors; Bottom: The output of $Q(\cdot)$.}\label{example2}
	\end{center}
\end{figure}

In the following, we show the ceiling effect caused by $R_1$. 
In the simulation example above, we have shown that $R_1=1$ is already sufficiently large, i.e., $\zeta_1/2^{R_1}< \rho(G)$. Thus, we verify the ceiling effect by an academic value $R_1=0.2$. According to Proposition \ref{theorem 1} c), one can choose $\gamma_1=0.97$. Similarly, we let $\gamma_2=1.10521$. Under this setting, we can see from the first plot of Figure \ref{example2} that tracking is achieved. Note that both $R_1$ and $\gamma_1$ now are smaller than the values in the previous examples, and hence the sufficient bound of tolerable DoS attacks is smaller. To recover the resilience to that in previous examples, we enlarge $\gamma_1$ for mitigating the bottleneck, i.e., $\gamma_1=0.93$ and $R_1=0.2$. However, as we presented in Proposition \ref{theorem 1} and (\ref{50}), only improving the bottleneck (beyond the ``ceiling") can cause overflow problems of follower's quantizers. This is shown in the second plot in Figure \ref{example2}, which implies that the variables to be quantized by $Q(\cdot)$ diverge. This implies that when $R_1$ is fixed and small, $R_1$ will determine the upper bound of resilience. One cannot further improve the resilience by violating the ceiling effect. Otherwise, the quantizer (\ref{quantizer}) will have saturation problems.

\begin{figure}[t]
	\begin{center}
				\includegraphics[width=0.35 \textwidth]{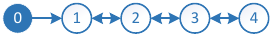} \\
				\vspace{2mm}
		\includegraphics[width=0.4 \textwidth]{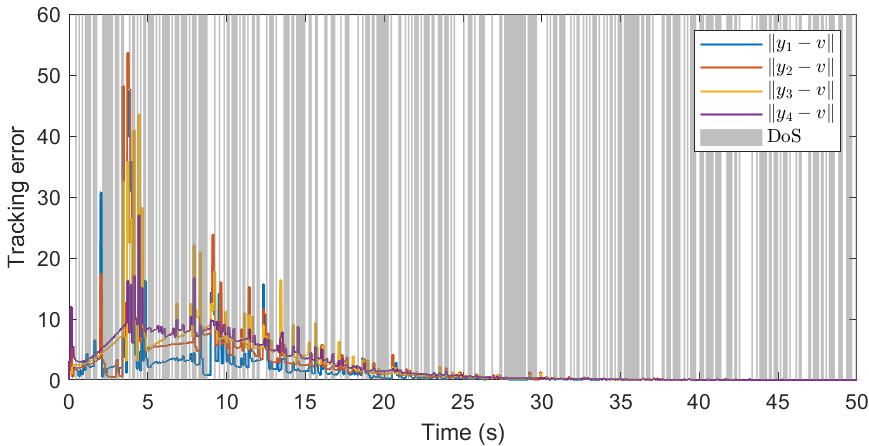} \\
				\includegraphics[width=0.4 \textwidth]{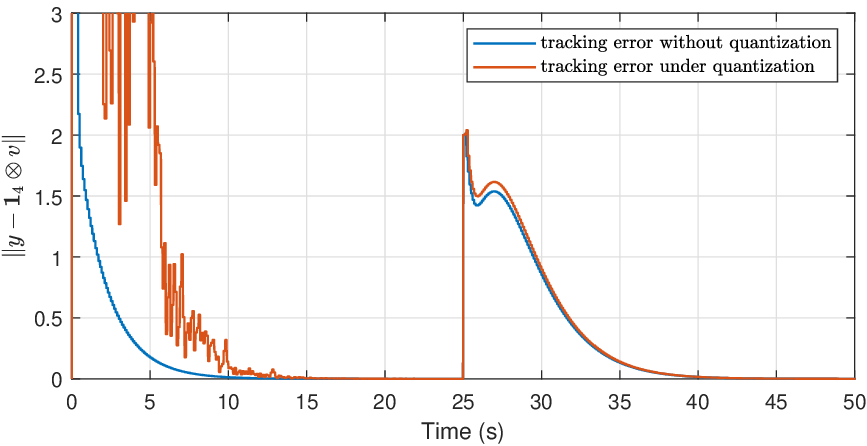} \\
		\vspace{-2mm}
		\linespread{1}\caption{Top: Communication topology with agent 0 being the leader and agents 1-4 being the followers; Middle: Time response of tracking errors under DoS; Bottom: Time response of tracking errors without and under quantization with $y:=[y_1 ^T \,\,y_2 ^T\,\,y_3 ^T\,\,y_4 ^T ]^T$.}\label{example3}
	\end{center}
\end{figure}

\textbf{Example 2:} 
In this example, we apply our control scheme to the tracking control of mobile robots under quantization and DoS attacks. The dynamics of four follower mobile robots is taken from \cite{guo2021linear} as 
$
\dot x_{i1}(t)= x_{i2}(t),\,\,\dot x _{i2}(t)= -c_ix_{i2}(t)+u_i(t),
$
in which $x_{i1}(t)$ and $x_{i2}(t)$ denote the position and velocity, respectively, with $i=1, 2, 3, 4$, and $c_i=0.2$ is the friction parameter. The virtual leader is an autonomous system with dynamics $\dot v(t) =[0\,\, 1;0 \,\,0]v(t)$ and the elements in $v(t)=[v_1(t) \,\,v_2(t)]^T$ denote the position and velocity. Let $y_i(t)=[x_{i1}(t)\,\, x_{i2}(t)]^T$ for $i=1, 2, 3, 4$. 
It is simple to obtain the system matrices for sampled-data dynamics with sampling interval $\Delta =0.1$s:
$A_i=[1.0000\,\,0.0990;0\,\,  0.9802]$, $B_i=[0.004967; 0.09901]$ for followers $i=1,2,3,4$ and $S=[1 \,\, 0.1;0\,\, 1]$ for the leader. The communication topology is illustrated in the top picture in Figure \ref{example3}. We select $K_i=[-77.2624\,\,-13.5990]$ and $\bar K=0.42I_2$. One can verify that $\rho(G)=0.9493$ and $\rho(S)=1$. Then we select $\gamma_1=0.95$, $\gamma_2=1.01$ and $R_1=1$ by Proposition \ref{pro 1}. By Theorem \ref{main result}, if $1/T+\Delta/\tau_D < 0.8375$, tracking control can be achieved. Similarly to Example 1, DoS attacks are generated randomly with $|\Xi(0.50)|=29.6$s and $n(0,50)=122$ as shown in the middle picture in Figure \ref{example3}, which yield $1/T+\Delta /\tau_D \approx 0.836$. It satisfies the bound of tolerable DoS attacks, i.e, $0.836 < 0.8375$. The time response of tracking errors are provided in the middle picture of Figure \ref{example3}, which presents the success of tracking control under DoS by $\|y_i-v\|\to 0$.  

We compare the performance of tracking control in the presence and absence of quantization. To make the comparison clear, DoS attacks are not considered. Moreover, we assume that the virtual leader increases its speed by 1m/s at $25$s and the followers should track the new speed as well as the position. As shown in the last plot in Figure \ref{example3}, the tracking errors converge to 0 in $[0, 25]$s and $(20, 50]$s with and without quantization. In particular, by the simulation curves in $(25, 50]$s, the convergence speed under quantization is close to the one without quantization. To further improve their convergence speeds, one needs to improve the communication topology, e.g., by unitizing the one in Example 1 and obtaining $\rho_{\text{new}}(G)=0.8304$. By the new $\rho_{\text{new}}(G)$, it is straightforward that the convergence speed without quantization is faster and the one under quantization is also faster since one can select a smaller $\gamma_1$.

\section{Conclusions and future research}

This paper studied tracking control of heterogeneous multi-agent systems under DoS attacks and state quantization. We have designed heterogeneous quantization mechanisms for the leader's and followers' state. With the dynamical quantized controllers with zooming-in and out capabilities, and proper data rate, we have shown that the quantizers are free of saturation under DoS attacks. 
Our results revealed the bottleneck effect of resilience: if one selects a ``large" data rate for leader-follower quantized communication, further enlarging it cannot improve the resilience. Then, the inter-follower quantized communication determines the resilience.  
We also presented the ceiling effect of resilience: by tuning the quantized controllers of followers, one can at most improve the resilience to the level determined by the data rate of leader-follower quantized communication. Otherwise, overflow of followers' state quantizer can occur.

For the directions of future research, it is interesting to consider the presence of noise. Then the dynamics of the tracking error is expected to be practically stable\cite{ran2021practical}. One can also design distributed scaling parameters such that each follower agent can have individual zooming-in and out factors. Moreover, it is useful to design controllers that can tolerate communication delays \cite{10075504}, uncertainties \cite{shariati2016descriptor} and accelerate consensus speed \cite{moradian2022study}. 

\section*{Appendix--Proof of Proposition \ref{pro 1}}

By Cases I--IV in Section 4, the evolution of $\alpha$ along with the sequence of successful steps $\{s_r\} \subseteq \{k\}$ follows: 
$
\alpha(s_{r})
= (\frac{\bar S_N }{\gamma_2})^{m_{r-1}}    \frac{G}{\gamma_1}\alpha(s_{r-1})     + ( \frac{\bar S _ N }{\gamma_2}\! )^{m_{r-1}}   
(\frac{P}{\gamma_1}  \xi_z (s_{r-1})  -   \frac{W}{\gamma_1} (\mathbf 1_N \otimes \xi_v(s_{r-1})) )
$, 
where $m_{r-1} \ge 0$ denote the number of unsuccessful transmission between $s_{r-1}$ and $s_r$.
There exists a unitary matrix $U$ such that 
$
U^T (\mathcal L_G + \mathcal D ) U
$ is an upper-triangular matrix whose diagonals are $\tilde \lambda _ i $. Then it is simple to obtain that
\begin{align}\label{32} 
&\bar G := (U \otimes I _v )^T G (U \otimes I _v ) = \nonumber \\ 
&\left[
\begin{array}{ccc}\renewcommand{\arraystretch}{0.1}
\bar S - \tilde   \lambda_1 \bar K  &\,\,\,\,\,\,\, * &  *   \\
&   \ddots &   *   \\
&   & \bar S - \tilde \lambda_N \bar K 
\end{array}
\right].
\end{align}
Recall that $\bar S$ is an upper-triangular matrix, then we assume that there exists an upper-triangular $\bar K$ such that $\bar S - \tilde \lambda_i \bar K$ is Schur stable for $i=1, \cdots, N$.

Define vectors $\bar \alpha(k) := (U \otimes I_{v}) ^T \alpha(k)$, $\bar \xi_z(k) := (U \otimes I_{v})^T \frac{P}{\gamma_1} \xi_z(k)$ and $ \bar \xi_v(k) := (U \otimes I_{v})^T \frac{W}{\gamma_1} (\mathbf 1 _N \otimes \xi_v(k))$. Then, one has   
$
\bar \alpha(s_{r})=(\frac{\bar S_N}{\gamma_2})^{m_{r-1}}    \frac{\bar G}{\gamma_1}    \bar \alpha(s_{r-1}) 
\,\, +  
(\frac{\bar S_N}{\gamma_2})^{m_{r-1}}  ( \bar \xi_z(s_{r-1})-\bar  \xi_v(s_{r-1}))
$, 
which is obtained by the fact that $(U \otimes I_v) ^T$ and $\bar S_N ^{m_r -1}$ are commuting matrices. 
Let $s_{-1}$ denote the step $k=0$. Then we have $\|\alpha(s_{-1})\| = \|\alpha(0)\| \le 2 \sqrt{N n_v}\frac{C_{x_0}}{\theta_0}$, $\|\bar  \xi_{z}(s_r)\|\le \|P\|\sqrt{N n_v}\frac{\sigma}{\gamma_1}$ and $\|\bar  \xi_{v}(s_r)\|\le \|W\|\sqrt{N n_v}\frac{E_v}{\gamma_1}$, and $\rho(\bar G/\gamma_1)<1$ and $\rho(\bar S_N/\gamma_2)< 1$ by the selections of $\gamma_1$ and $\gamma_2$. By the stabilization of switched systems, there exists a positive real $C_1$ such that $\|\bar \alpha (s_r) \|  \le C_1$.  
In particular, $\bar S _N $ and $\bar G $ are upper-triangular, which implies there are no dwell-time constraints.


To analyze the overflow problem, i.e., the lower bound of $(2R_f +1) \sigma$, one needs to investigate $\|\frac{\bar z(k) - \bar S_N \hat z(k-1)}{\theta_{k-1}}\|_\infty$ by (\ref{eq estimator}). At $s_r + 1$, one has 
$
\left\|(\bar z(s_r + 1) - \bar S_N \hat z(s_r))/\theta_{s_r}\right\|_\infty 
=\| Z \xi_z(s_r)  -  P \alpha(s_r)   - W (\mathbf 1_N \otimes \xi_v(s_r)) \|_\infty 
 \le \| Z \xi_z(s_r)  -  P \alpha(s_r)   - W (\mathbf 1_N \otimes \xi_v(s_r)) \|  
\le (2R_f+1)\sigma
$
which implies unsaturation of follows' quantizer $q_{R_f}(\cdot)$. 
For the step of $s_r + 2$, if $s_r+1$ is a successful transmission step, one can follow the similar analysis presented in the step of $s_r + 1$ and shows quantizer unsaturation. If $s_r+1$ is not a successful transmission step, then one has 
$
\left\| \frac{\bar z(s_r + 2) - \bar S_N \hat z(s_r+1)}{\theta_{s_r+1}}\right\|_\infty 
=\| \bar S _N \xi_z (s_r+1) \|_\infty \le (2R_f+1)\sigma
$,
in which we have used $\|\xi_z(s_r + 1 )\|_\infty \le \| Z \xi_z(s_r)  -  P \alpha(s_r)   - W (\mathbf 1_N \otimes \xi_v(s_r)) \|_\infty $. By induction, if the transmissions at $s_r + m-1$ are all failed, then for $s_r + m$ ($m\ge2$), one has 
$
\left\|\frac{\bar z(s_r + m) - \bar S_N \hat z(s_r+m-1)}{\theta_{s_r+m-1}}\right\|_\infty 
=\| \bar S _N \xi_z (s_r + m-1) \|_\infty 
=\| \bar S _N \left(\frac{\bar S_N}{\gamma_2}\right)^{m-2} \xi_z (s_r +1) \|_\infty  \le (2R_f+1)\sigma
$,
in which $\rho( \bar S_N/\gamma_2)< 1$ such that $\|(\bar S_N /\gamma_2)^{k}\| \le C_2 \in \mathbb R _{>0}$. By the above analysis, one can see that the quantizer $q_{R_f}(\cdot)$ does not saturate at $s_r$, $s_r + 1$ and $s_r+m$ ($m\ge 2$), which implies quantizer unsaturation at all $k$.
\qedp

\vspace{-1mm}

\small

\bibliographystyle{unsrt}
\bibliography{2}

\end{document}